\documentclass[12pt]{article}
\begin{document}
\title{Ultra High Energy Decaying Fermions}
\author{B.G. Sidharth\\
International Institute for Applicable Mathematics \& Information Sciences\\
Hyderabad (India) \& Udine (Italy)\\
B.M. Birla Science Centre, Adarsh Nagar, Hyderabad - 500 063
(India)}
\date{}
\maketitle

\section{Introduction}
The LHC in Geneva is already operating at a total energy of $7 TeV$
and hopefully after a pause in 2012, it will attain its full
capacity of $14 TeV$ in 2013. These are the highest energies
achieved todate in any accelerator. It is against this backdrop that
it is worthwhile to revisit very high energy collisions of Fermions
(Cf. also \cite{bgspp}). We will in fact examine their behaviour at
such energies.
\section{The High Energy Equation}
It is known that at very high energies, we encounter negative
energies. This is because the set of positive energy solutions of
the Dirac or Klein-Gordon equations is not a complete set
\cite{feshbach}. At usual energies we could apply the well known
Foldy-Wothyson transformation to recover a description in terms of
positive energies alone or more precisely a description free of
operators which mix negative energy and positive energy components
of the wave function. This description also leads in the
non-relativistic limit to the two component Pauli equation
\cite{bd}.\\
In the case of very high energies it was shown several years ago by
Cini and Toushuk that we can modify the Foldy-Wothyson
transformation and obtain a different description \cite{cini}. Let
us examine this situation in
greater detail \cite{schweber}.\\
The Cini-Toushuk transformation can be written in the form
\begin{equation}
e^{\pm \imath s} = \frac{E + |p|}{2E} \pm \frac{\vec{\gamma} \cdot
\vec{p}}{2E|p|} \cdot m\label{I}
\end{equation}
Under (\ref{I}), it is well known that the Dirac equation takes on
the form of the massless neutrino equation:
$$H \psi = \frac{\vec{\alpha} \cdot \vec{p}}{|p|} E(p)$$
In the above we use the following notation:
\begin{equation}
\alpha^k = \left(\begin{array}{ll} 0 \quad \sigma^k\\
\sigma^k \quad 0\end{array}\right) \quad \quad \beta = \left(\begin{array}{ll} I \quad 0\\
0 \quad -I\end{array}\right)\label{24}
\end{equation}
\begin{equation}
\gamma^0 = \beta\label{26}
\end{equation}
\begin{equation}
\gamma^k = \beta \alpha^k \quad (k = 1,2,3)\label{27}
\end{equation}
where $\alpha^k$ are the Pauli matrices and $I$ is the $2 \times 2$
unit matrix.\\
We will also require the transformation of the $\gamma_5$ operator,
which is, given by,
\begin{equation}
\gamma_5 = \gamma^0 \gamma^1 \gamma^2 \gamma^3 = \imath \left(\begin{array}{ll} I \quad 0\\
0 \quad -I\end{array}\right)\label{86b}
\end{equation}
Using (\ref{I}), the transformed matrix is given by,
\begin{equation}
\Gamma_5 = e^{- \imath s} \gamma_5 e^{\imath s} = \left\{ \frac{E +
p}{2E} + \frac{(\vec{\gamma} \cdot \vec{n})m} {2E} \right\} \gamma_5
\left\{ \frac{E+p}{2E} - \frac{(\vec{\gamma} \cdot \vec{n})m} {2E}
\right\}\label{II}
\end{equation}
which finally reduced to
\begin{equation}
\Gamma_5 = \gamma_5 + \left(\frac{m}{E}\right) \left(\vec{\gamma}
\cdot \vec{n}\right) \gamma_5\label{III}
\end{equation}
In the above $\vec{n}$ is the unit vector in the direction of the
momentum vector. We can see from (\ref{III}) that
\begin{equation}
\Gamma_5 = \gamma_5\label{IV}
\end{equation}
whenever $m$ vanishes. This is of course the well known two
component neutrino case where the wave function can be decomposed
into the left handed and right handed neutrino wave functions. Let
us use (\ref{III}) to proceed along similar lines and write
\begin{equation}
\psi = \psi_1 + \psi_2 \cdots\label{V}
\end{equation}
where
\begin{equation}
\psi_1 = \frac{1}{2} (1 - \gamma_5) \psi \, \mbox{and}\, \psi_2 =
\frac{1}{2} (1 + \gamma_5) \psi\label{VI}
\end{equation}
If (\ref{IV}) were to hold, as in the neutrinos, then (\ref{VI})
would be the decomposition in terms of the left handed and right
handed wave functions. If the mass does not vanish, that is
(\ref{IV}) does not hold then we will have from (\ref{VI})
\begin{equation}
\psi_1 = (1 + \frac{m}{E}) (1 + \gamma_5) \psi - \frac{m}{E} \psi =
(1 + \frac{m}{E}) \psi_L - \frac{m}{E} \psi\label{VII}
\end{equation}
with a similar equation for $\psi_2$. Equations (\ref{V}) and
(\ref{VII}) show that if $\frac{m}{E}$ is much less than $1$, that
is when the energy is much greater than the rest energy, then we
have a nearly two component neutrino like situation. We could for
example interpret (\ref{V}) and (\ref{VI}) as a decomposition into
the left and right handed wave functions where the particle, as can
be seen from (\ref{VII}) nearly exhibits handedness. Or more
specifically as can be seen from (\ref{VII}) the wave function has a
large part that displays handedness and a small part which is the
usual type of wave function. More generally we can write (\ref{VII})
as
\begin{equation}
\psi = \psi_H + \alpha \psi_\Delta\label{VIII}
\end{equation}
where $\psi_{\hat{H}}$ is the handed part and the second term is a
small correction.\\
It must be borne in mind that when the total energy is much greater
than the rest energy (\ref{VIII}) holds. One could hope to see the
effects, hopefully in the LHC which as remarked has already reached
the $7 TeV$ mark and is expected to reach the $14 TeV$ mark sometime
in 2013.
\section{Possible Consequences}
Firstly, it must be observed that the above theory becomes relevant
in view of the fact that the neutrino is now known to have a mass,
though the mass values are not yet certain, unlike the mass
differences. This is because equations like (\ref{VI}), (\ref{VII})
and (\ref{VIII}) can now be applied to neutrinos. This apart the
above shows that Fermions in general behave like "heavy" neutrinos
at very high energies. In any case as can be seen, these equations
imply that apart from a $O(\frac{m}{E})$ correction, $\gamma_5$ gets
multiplied effectively by a factor $(1 + O(\frac{m}{E}))$
(Cf.(\ref{VII})). This means that in the usual Salam-Weinberg theory
a typical interaction term gets multiplied by a factor $(1 + O
(\frac{m}{E}))$ \cite{taylor}
\begin{equation}
2^{\frac{1}{2}} G_w \left\{\bar{\nu}_\mu \gamma^\lambda \frac{1}{2}
(1 + \gamma_5) \nu_\mu\right\} \left\{\bar{e} \gamma_\lambda
\left[\frac{1}{2} (1 + \gamma_5)c_L + \frac{1}{2} (1 - \gamma_5)c_R
\right]e\right\} (1 + O(\frac{m}{E})),\label{8.41}
\end{equation}
That is $c_L$ and $c_R$ are also multiplied by a similar small
deviation from unity to become $c'_L, c'_R$. This in turn implies
that the differential cross section now becomes, in terms of the
fermion recoil energy $E'$
$$\frac{d\sigma}{dE'_0} = [G^2_w/(2\pi m_eE^2_\nu)] [|c'_L|^2 (p
\cdot q)^2 + |c'_R|^2 (p' \cdot q)^2$$
\begin{equation}
\quad \quad + \frac{1}{2} (c'^{*}_R c'_L + c'^{*}_Lc'_R) m^2_0 q
\cdot q'].\label{8.42}
\end{equation}
In any case the use of $\Gamma_5$ given by (\ref{III}) instead of
$\gamma_5$ would mean that a decay process would be asymmetrical in
the angular distribution of the type $(1 + P cos \Theta$) where $P$
is the average polarization.\\
The point is that fermions at such high energies would show
handedness in accordance with (\ref{VII}) or (\ref{VIII}). The
possibility of CP violation in ultra high energy cosmic rays has
been discussed by Collady and others \cite{collady}. In any case,
these effects would have been present in the early universe.\\
Sudarshan et al \cite{sudarshan} use a similar analysis to get
positive and negative energy operators $\chi \pm$ for position and
similar momentum operators, but  interestingly they show that the
$x$ and $y$ components do not commute. Sudarshan and co-workers
introduced a sub or superscript $D$ and $E$ for the Dirac and
extreme relativistic (that is Cini-Toushek type) representations.
Then they deduced
$$[x_\pm , y_\pm ] = \left(\frac{\imath p_2}{2 p^3} \gamma_5
\Lambda_\pm E\right)_{E \, repres.}$$
\begin{equation}
\quad \quad = \left(\pm \frac{\sigma_2}{2 \imath p^2} \Lambda_\pm
D\right)_{D \, repres.}\label{11s}
\end{equation}
where $\Lambda$ is a projection operator which is given by
$$\Lambda_\pm = \frac{1}{2} (1 \pm H/E)$$
in this representation. This matter was investigated by Newton and
Wigner \cite{newton} from a slightly different angle. Some years ago
the author revisited this aspect from yet another point of view
\cite{bgsfplextn} and showed that this noncommutativity which is
exhibited by (\ref{11s}) is related to spin and extension. The
noncommutative nature of spacetime has been a matter of renewed
interest in recent years. At very high energies, it has been argued
that \cite{refs} there is a minimum fuzzy interval, symptomatic of a
non commutative spacetime, so the usual energy momentum relation
gets modified and becomes
\begin{equation}
E^2 = p^2 + m^2 + \alpha l^2 p^4\label{a}
\end{equation}
the so called Snyder-Sidharth Hamiltonian \cite{bgs1,bgs2,bgs3}.
Using (\ref{a}) it is possible to deduce the ultrarelativistic Dirac
equation \cite{uhedeijmpe}
\begin{equation}
(D + \beta l p^2 \gamma^5 ) \psi = 0\label{b}
\end{equation}
$\beta = \sqrt{\alpha}$. In (\ref{b}) $D$ is the usual Dirac
operator above while the extra term appears due to the new
dispersion relation (\ref{a}). We can see from (\ref{b}) that the
Hamiltonian now becomes non Hermitian and takes on an extra term:
\begin{equation}
H = M -\imath N\label{c}
\end{equation}
where $M$ is the usual Hamiltonian and $N$ is now Hermitian
(Cf.\cite{neutrino}), that is, $M$ and $N$ are real. This indicates
a decay. With the modified Dirac equation (\ref{b}) in place of the
usual Dirac equation, we can now treat the two states considered
above viz.,
$$\psi_L , \psi_R$$
as forming a two state system in this subspace of the Hilbert space
of all states where the two components decay at different rates, in
general as we will see below. The theory of such two state systems
is well known \cite{feynmanlectures}.  In fact the two states would
now be given by
\begin{equation}
\psi_{L,R}(t) = e^{\imath Mt} \cdot e^{-Nt} \psi_{L,R}(0)\label{d}
\end{equation}
where the left side refers to the sate of time $t$ and the right
side wave function to the time $t = 0$ (Cf.\cite{widhall}). We can
write the Hamiltonian (\ref{c}) above for the two state as
$$H_{eff} = \left(\begin{array}{ll}
H_{11} \quad H_{12}\\
H_{21} \quad H_{22} \end{array}\right) = M - \frac{\imath}{2} N =
\left(\begin{array}{ll}
M_{11} \quad M_{12}\\
M_{21} \quad M_{22}\end{array}\right) - \frac{\imath}{2}
\left(\begin{array}{ll}
N_{11} \quad N_{12}\\
N_{21} \quad N_{22}\end{array}\right)$$ where-by virtue of the
pulled out $\imath$-both $M$ and $N$ are Hermitian. An additional
constraint, namely $H_{11} = H_{22}$, comes from the CPT theorem.
Let us continue with the
two state analysis.\\
The evolution equation (in this 2-D approach),
$$H|\psi> = \imath \frac{d}{dt}| \psi >$$
yields the usual solution
$$|\psi_{H,L} > (t) = exp^{-\imath H_{H,L}} | \psi_{H,L} > (0)$$
where $H_{H,L}$ denotes the eigenvalues of $H$, which are under the
assumption of CPT symmetry given as is well known, by
$$H_{H,L} = H_{11} \pm \sqrt{H_{12} H_{21}}$$
and $|\psi_{H,L}>$ are eigenstates of the form
$$|\psi_{H,L}> = p|\psi^0 > \mp q|\bar{\psi}^0 >$$
with
$$\frac{q}{p} = - \frac{H_H - H_L}{2H_{12}}$$
(Cf.ref.\cite{widhall}). Rewriting the time-dependent solution using
$H_{H,L} = M_{H,L} - \frac{\imath}{2} N_{H,L}$ with real $M$ and
$N$, we get
$$|\psi_{H,L}> (t) = exp \left[- \frac{N_{H,L}}{2} \right]exp^[{-\imath
M_{H,L}^t]} |\psi_{H,L}> (0)$$ This represents two Fermions (one
perhaps heavier with mass $M_H$, one Lighter with mass $M_L$),
decaying with (generall different) decay constants $N_{H,L}$. The
mean mass $M = \frac{1}{2} (M_H + M_L) \, \mbox{and} \, \Delta M: =
M_H - M_L$. It has been pointed out \cite{bgsarxiv} that equations
like (\ref{VI}), (\ref{VII}) or (\ref{VIII}) applied to neutrinos
which are massless suggests one (or more) neutrinos. This is brought
out more clearly in the above. Remarkably there seems to be very
recent confirmation
of such an extra or sterile neutrino \cite{roe}. Here a particle that is approaching the speed of light, ultimately decays, something which may be relevant for the LHC. Theoretically as the velocity of light is approached the particle acquires infinite momentum and energy -- the decay prevents this. It would be interesting to investigate if in such a decay, any laws like CP are violated. Indeed this seems to be the case for $B$ and $K$ decays.\\
In any case this analysis is true for Fermions in general, one would
expect handedness and even decomposition at very high energies. One
could look at it in the following way. The new Hamiltonian
(\ref{a}), the modified Dirac equation (\ref{b}) and the non
Hermition Hamiltonian (\ref{c}) split the st ate, much like the
introduction of a magnetic field leading to the Zeeman splitting.
\newpage
\begin{flushleft}
{\bf {\large {Appendix}}}
\end{flushleft}
\noindent It is interesting that in the theory of Bosons too, we
encounter a situation similar to that discussed above, with two
states and a non Hermition Hamiltonian. That is because in Quantum
Mechanics we encounter negative energies, unlike in Classical
Physics. In the case of the Dirac electron, this lead to the
postulation of the Hole theory. Let us now start with the
Klein-Gordon equation. As has been shown in detail by Feshbach and
Villars \cite{feshbach}, we can rewrite the K-G equation in the
Schrodinger form, invoking a two component wave function,
\begin{equation}
\Psi = \left(\begin{array}{ll} \phi \\
\chi\end{array}\right),\label{2.16} \end{equation} The equation is
$$\imath \hbar (\partial \phi /\partial t) = (1/2m) (\hbar /\imath
\nabla - eA/c)^2 (\phi + \chi)$$
$$\quad \quad \quad +(e\phi + mc^2)\phi,$$
\begin{equation}
\imath \hbar (\partial \chi / \partial t) = - (1/2m) (\hbar / \imath
\nabla - eA/c)^2 (\phi + \chi) + (e\phi - mc^2)\phi\label{2.15}
\end{equation}
It will be seen that the components $\phi$ and $\chi$ are coupled in
(\ref{2.15}). In fact we can analyse this matter further,
considering free particle solutions for simplicity. We write, $$\Psi
= \left(\begin{array}{ll} \phi_0 (p) \\ \chi_0 (p)\end{array}\right)
\, e^{\imath / \hbar (p\cdot x-Et)}$$
\begin{equation}
\Psi = \Psi_0 (p)e^{\imath / \hbar (p\cdot x-Et)}\label{2.25}
\end{equation}
Introducing (\ref{2.25}) into (\ref{2.15}) we obtain, two possible
values for the energy $E$, viz.,
\begin{equation}
E = \pm E_p ; \quad E_p = [(cp)^2 +
(mc^2)^2]^{\frac{1}{2}}\label{2.26}
\end{equation}
The associated solutions are
$$\left.\begin{array}{ll} E = E_p  \quad \phi_0^{(+)} =
\frac{E_p+mc^2}{2(mc^2E_p)^{\frac{1}{2}}}\\
\psi_0^{(+)}(p): \quad \chi_0^{(+)} =
\frac{mc^2-E_p}{2(mc^2E_p)^{\frac{1}{2}}}\end{array}\right\}\phi_0^2
- \chi_0^2 = 1,$$
\begin{equation}
\left.\begin{array}{ll} E = - E_p  \quad \phi_0^{(-)} =
\frac{mc^2 - E_p}{2(mc^2E_p)^{\frac{1}{2}}}\\
\psi_0^{(-)}(p): \quad \chi_0^{(-)} = \frac{E_p +
mc^2}{2(mc^2E_p)^{\frac{1}{2}}}\end{array}\right\} \phi_0^2 -
\chi_0^2 = -1\label{2.27}
\end{equation}
It can be seen from this that even if we take the positive sign for
the energy in (\ref{2.26}), the $\phi$ and $\chi$ components get
interchanged with a sign change for the energy. Furthermore we can
easily show from this that in the non relativistic limit, the $\chi$
component is suppressed by order $(p / mc)^2$ compared to the $\phi$
component exactly as in the case of the Dirac equation \cite{bd}.
Let us investigate this circumstance further. In (\ref{2.15}) if we
take
\begin{equation}
\frac{1}{2m} (-\imath \nabla - eA)^2 \chi = 0\label{wA}
\end{equation}
then we have
\begin{equation}
\imath \dot{\phi} = \frac{1}{2m} (-\imath \nabla - eA)^2 \phi +
(eA_0 + mc^2)\phi\label{wB}\end{equation} and also $$\imath
\dot{\chi} = -\frac{1}{2m} (-\imath \vec{\nabla} -cA)^2 \phi +
(eA_0-mc^2)\chi$$
\begin{equation}
= -\imath \dot{\phi} + (eA_0 + mc^2) \phi + (eA_0 - mc^2)
\chi\label{wC}
\end{equation}
It can be seen that (\ref{wA}) and (\ref{wB}) are Schrodinger
equations and so solvable. However (\ref{wC}) couples $\phi$ and
$\chi$. In fact we have
\begin{equation}
\dot{\phi} + \dot{\chi} = (eA_0 + mc^2) (\phi + \chi) - 2mc^2
\chi\label{wD}
\end{equation}
In the case if
\begin{equation}
mc^2 > > eA_0 \quad (\mbox{or} \, A_0 = 0)\label{wF}
\end{equation}
we can easily verify that
\begin{equation}
\phi = e^{\imath px-Et} \mbox{and} \, \chi = e^{\imath
px+Et}\label{wG}
\end{equation}
is a solution.\\
That is $\phi$ and $\chi$ belong to opposite values of $E (m \ne
0)$. The above shows that K-G equation mixes the positive and
negative energy solutions. If on the other hand $|mc^2| <<1$, then
(\ref{2.27}) shows that $\chi$ and $\phi$ are of same energy that is
$m_0 c^2 \approx 0$ that is $m_0 = 0$. This shows that if $\phi$ and
$\chi$ both have the same sign for $E$, that is there is no mixing
of positive and negative energy, then the rest mass $m_0$ vanishes.
Further we are now in a position to argue that solutions with a
single sign of the energy have no rest mass. A non vanishing rest
mass requires the mixing of both signs of energy. Indeed it is a
well known fact that for solutions which are localized about a point
$x_0$ in the $\delta$ function sense, both signs of the energy
solutions are required to be superposed \cite{schweber}. This is
because only positive energy solutions or only negative energy
solutions do not form a complete set. Interestingly the same is true
for localization
about a time instant $t_0$.\\
Further, we observe that
\begin{equation}
t \to -t \Rightarrow E \to - E, \quad \phi \leftrightarrow
\chi\label{wJ}
\end{equation}
Let us write (\ref{wC}) as `(with $\hbar = 1 = c)$
\begin{equation}
H \phi = H_{11} \phi + H_{12} \chi\label{waee}
\end{equation}
and similarly we have
\begin{equation}
H \chi = H_{21} \phi + H_{22} \chi\label{wbee}
\end{equation}
We now observe that in Quantum Field Theory, a sub space of the full
Hilbert can exhibit the complex or non Hermitian Hamiltonian of the
type encountered above.\\
Writing $H = M - \imath N$ as before we have
$$M_{11} = - \frac{\hbar^2}{2m} \nabla^2 + \frac{1}{2m}
\frac{e^2A^2}{c^2} + (e \phi + mc^2)$$
$$M_{21} = + \frac{\hbar^2}{2m} \nabla^2 - \frac{e^2A^2}{c^2} + (e
\phi - mc^2)$$
$$N_{11} = \frac{1}{m} \frac{eA}{c} \hbar \nabla = N_{12}$$
\begin{equation}
N_{21} = - N_{11} = N_{22}\label{wcee}
\end{equation}
We can now treat $|\phi , \chi >$ as a two state system and further
it follows from the above that
$$|\phi , \chi > (t) = exp (- N_{12} t) exp.$$
\begin{equation}
\quad \quad exp \left(- \imath M_{12} t\right) | \phi , \chi >
(0)\label{wdee}
\end{equation}
Equation (\ref{wdee}) shows that the states $|\phi >$ and $| \chi >$
decay, but decay at different rates.\\
Treating $| \phi >$ and $| \chi >$ as particle and anti particle, we
have exactly this situation in $B$ and $K^0$ decay. The point here
is that as in the case of the $B$ or $K^0$ mesons, the decay rates
of the particles and antiparticles would be different, thus leading
to a CPT violation. The above considerations provide an explanation.
A full discussion will be given later.
\begin{flushleft}
{\bf {\large {Remarks}}}
\end{flushleft}
\noindent i) From the above analysis it is clear that a localized
particle requires both signs of energy. At relatively low energies,
the positive energy solutions predominate as per particles as very
high energies. On the other hand it is the negative energy solutions
that predominate as for the negatively charged counterpart or the
anti particles.\\
ii) In other words write the wave function as
$$\psi = \psi_+ + \psi_-$$
iii) We also observe that the symmetry (\ref{wJ}) holds good. On the
other hand we have shown in detail that the Schrodinger equation
goes over to the Klein-Gordon equation if we allow $t$ to move
forward and also backward. Here we have done the reverse of getting
the Klein-Gordon equation into two Schrodinger equations. This is
expressed by (\ref{wJ}). From here we can argue that the Hamiltonian
$H$ for the time reversed system becomes imaginary compared to the
usual time system.\\
In any case we would like to stress that the two degrees of freedom
associated with the second time derivative can be interpreted,
following Pauli and Weisskopf as positive and negatively charged
particles or particles and anti particles.\\
iv) Let us remain in the realm of point particles. The point is that
if we approach distances of the order of the Compton wavelength, the
negative energy solutions begin to dominate, and we encounter the
well known phenomenon of Zitterbewegung. This modifies the coupling
of the positive solutions with an external field, particularly if
the field varies rapidly over the Compton wavelength. In fact this
is the origin of the well known Darwin term in the Dirac theory
\cite{bd}. The Darwin term is a correction to the interaction of the
order
\begin{equation}
\left(\frac{p}{mc}\right)^4 \, \mbox{and} \,
\left(\frac{p}{mc}\right)^2\label{13}
\end{equation}
for spin $0$ and spin $1/2$ particles respectively.\\
v) We also finally observe that in the case of a particle with
vanishing mass, the positive and negative energy solutions decouple.
In other words there is no localization in the particle sense.\\
vi) Cf. also p.33 (f-v): Only positive energy solutions can form a
classical type particle but with  extension of Compton wavelength --
not a point particle. It is a particle $> >$ Compton wavelength. At
Compton wavelength we begin to see negative energy solutions. These
modify the coupling of positive amplitudes to field (SS).\\
vii) In SS, $E^2 = p^2 + m^2 - \alpha l^2 p^5$, of $p > > 1$, then
$E^2 \alpha 0$ and $E$ becomes imaginary! True of $\alpha l^2 p^4 >
p^2 + m^2$ ; if $p^2 > m^2$, then $(\alpha \sim 1) \Rightarrow p^2
\frac{1}{m^2} > 1 (l = \frac{1}{m}) \, \mbox{or} \, p^2 > m^2 \to$
true. All this happens when $O(l^2) \ne 0$ that is Noncommutative
Geometry.

\end{document}